\documentclass[preprint,showpacs,preprintnumbers,amsmath,amssymb]{revtex4}

\usepackage{graphicx}
\usepackage{dcolumn}
\usepackage{bm}

\begin{document}

\preprint{APS/123-QED}

\title{Electronic structure and magnetic properties of CrSb$_2$ and
  FeSb$_2$ investigated via ab-initio calculations }

\author{G. Kuhn}
\author{S.~Mankovsky}
\author{H. Ebert}
\affiliation{
University of Munich,  
Department of Chemistry, 
Butenandtstrasse 5-13, D-81377 Munich, Germany
}%
\author{M. Regus}
\author{W. Bensch}
\affiliation{Institut f\"ur Anorganische Chemie, Christian-Albrechts-Universit\"at zu Kiel,
Max-Eyth-Str. 2, D-24118 Kiel,~Germany
}%

\date{\today}

\begin{abstract}

The electronic structure and magnetic properties of CrSb$_2$ have been
investigated by ab-initio calculations with an emphasis on the role of
the magnetic structure for the ground state. The
influence of correlation effects has been investigated  by performing
fixed spin moment (FSM) calculations showing their important role
for the electronic and magnetic properties. 
The details of the electronic structure of CrSb$_2$ are analyzed by a 
comparison with those of FeSb$_2$. The results obtained contribute in
particular to the understanding of the temperature dependence of
transport and magnetic behavior observed experimentally. 

\end{abstract}

\pacs{Valid PACS appear here}
\maketitle

\section{\label{sec:level1} Introduction }

According to the experimental data available in the literature, the
CrSb phase diagram shows two stable phases, having NiAs (CrSb) and
marcasite (CrSb$_2$) structures \cite{HK68,HK70,BK70}, respectively.
In the case of CrSb$_2$ with marcasite structure (space crystal group  Pnnm) 
the Cr atoms are octahedrally  coordinated by six nearest neighbor Sb
atoms.  As discussed in the literature, e.g. \cite{HM65}, CrSb$_2$ belongs
together with FeSb$_2$ to the  so-called 'Jahn-Teller' or 'regular' class of
marcasites (characterized by a ratio of the lattice parameters $c/a \approx 0.53-0.57$, $c/b \approx 0.48$) in
contrast to 'anomalous' marcasites (characterized by $c/a \approx 0.73-0.75$,
$c/b \approx 0.62$)\cite{HM65}. The
'regular' marcasites have an electronic structure 
leading to a Jahn-Teller instability  with respect to octahedral
symmetry, i.e. to a compression of the octahedral surrounding. This
leads to a $C_{2h}$ symmetry at the transition metal (TM) sites that gives rise to a
corresponding splitting of the $d$-states in the crystal field
\cite{HM65,Goo72}. According to the element specific occupation of
the splitted d-shell one can distinguish between the properties of low-spin
$3d^4$ configuration occurring for FeSb$_2$ and  high-spin $3d^2$
configuration of CrSb$_2$.  Although  the CrSb$_2$ and FeSb$_2$
compounds  differ concerning these features, 
they have 
very similar crystal structures and therefore, it seems worth 
to consider them in parallel. In particular, is also motivated by the
fact that only little information on CrSb$_2$ is available while
the properties of FeSb$_2$ are investigated experimentally and
theoretically in an extensive way \cite{PLV+05,ZSG+11,SSIS11,LMA+06,KA08,MHF+12}.
From the literature it is known that FeSb$_2$ and CrSb$_2$ compounds are
narrow-gap semiconductors at ambient pressure \cite{HK68,HK70,BK70}.
While FeSb$_2$ is nonmagnetic (NM) at low temperature,
CrSb$_2$  has an antiferromagnetic (AFM) order with a magnetic moment of the Cr
atom of about $1.94 \mu_B$ \cite{HKA70,HMP07}. 
A temperature increase results in a transition from the small gap
semiconductor to a metallic state with strong spin moment fluctuations
\cite{PLV+05}. This behavior is attributed to strong electron-electron
correlation effects \cite{PLV+05,LMA+06,KA08,MSL+10,SSIS11} that gives
rise to the properties observed experimentally: the temperature
dependence of magnetic susceptibility \cite{SSIS11}, the temperature
dependent behavior of the electrical resistivity $\rho(T)$
\cite{BJM+07,SOJ+09} and  
colossal thermopower $S(T)$ at $T = 10$ K  \cite{BJM+07,SOJ+09}.
This behavior was found to be similar to that observed
in FeSi with a Kondo insulator (KI) model as a possible scenario to describe
the observed properties which have been investigated theoretically via
LDA+U electronic structure calculations \cite {AEE+96}. The
corresponding investigations on FeSb$_2$ have been performed by
Lukoyanov et al. \cite{LMA+06} giving arguments for the KI
description of the spin state in FeSb$_2$. 
However, the origin of the electronic properties of FeSb$_2$ as well as
of FeSi, is still under discussion. In particular the recent theoretical
investigations \cite{KA08,MSL+10} gave  arguments against the KI scenario
of transition in these systems and discuss the effect of strong local
dynamical correlations. 

As was mentioned above, CrSb$_2$ has an AFM order below 
 $T \approx 273$ K with a local magnetic moment of Cr atom of
about $1.94 \mu_B$ \cite{HKA70,HMP07}.
The heat capacity observed in experiment has a lambda-shaped maximum at
$T_N = 274.08$ K \cite{AFE+78}, which corresponds to a transition to the
AFM state upon temperature decrease.
The magnetic susceptibility $\chi(T)$ has an anomaly at $T = 55$ K which 
may be attributed \cite{THK+08}  to a crossover between the different electronic
states in the AFM semiconductor, but cannot be related to any magnetic
transformation. No anomaly of the magnetic susceptibility has been observed
near $T_N =  273$ K \cite{THK+08}. On the other hand, the electrical resistivity
 measurements show an anomaly at the Neel temperature $273$ K,
and at the same time has a plateau in the temperature regime $50-80$ K. 
This finding is supposed to have a common origin with the susceptibility anomaly
and is also attributed to a transition between the different
semiconductive  electronic  states with the energy gap evaluated as $\Delta
= 0.07$ eV.  The behavior sketched for the resistivity has been
 observed by various experimental groups
 \cite{HKT+04,HMP07,THK+08,LQLX09}.
In addition, Hu et al. \cite{HMP07} have demonstrated a rather pronounced anisotropy in
the transport properties of FeSb$_2$ which is not observed in CrSb$_2$.
The measurements of the transverse magnetoresistance (TM)
$\frac{\rho(H) - \rho(0)}{\rho(0)}$ in a magnetic field up to $H = 90$
kOe at different temperatures exhibits an increase of this value  with
$H^2$ and then linearly as $H$ increases further \cite{THK+08}. As has
been pointed out \cite{THK+08}, a similar
behavior of the TM has been observed also for the FeSi semiconductor.
The temperature dependent behavior of $\frac{\Delta\rho}{\rho(0)}$
exhibit a maximum at around  $\approx 20K$.

Li et al. \cite{LQLX09} have shown that the thermal conductivity of
CrSb$_2$ has a maximum at $T \approx 50$ K, that they attributed
to the effect of phonon-phonon Umklapp scattering.
The thermopower $S$  measured as a function of temperature 
is negative in the whole temperature range and has a large
peak at $\sim 60$ K which correlates with the position of the plateau 
of the temperature dependent resistivity $\rho(T)$. For comparison, the
thermopower $|S|$ in FeSb$_2$ 
has a giant maximum at $T \approx 10K$ and its thermal conductivity
at $\approx 15$ K. \cite{BJM+07,SOJ+09}

A high-pressure experiment on CrSb$_2$ showed that for a pressure above $5.5$ GPa
a phase transition  to the CuAl$_2$ structure occurs accompanied  by 
 ferromagnetic (FM) order with a Curie temperature $T \approx 160$ K and a saturation magnetic
moment of about $1.2 \mu_B$ per Cr atom  \cite{TUE99}.
Experiments on FeSb$_2$ show no phase transitions 
up to $7$ GPa \cite{PLV+05}. This is in line with
theoretical investigations that predict that for a transition from the marcasite to the
CuAl$_2$  phase the pressure of about $41$ GPa is required \cite{WSQ+09}.

\section{Computational details}

The calculation are done with the ELK code based on the FP-LAPW method
\cite{ELK}. To 
take into account correlation effects for Fe and Cr the calculations
were done within the GGA as well as the GGA+U method. We used the exchange
correlation functional by Perdew, Burke, Ernzerhof (PBE)
\cite{PBE96}. For the GGA+U calculations the Coulomb ($U$) and exchange
($J$) parameters were chosen  as 
follows: $U = 2.7$ eV and $J = 0.3$ eV for Cr and $U = 2.6$ eV and
$J = 0.88$ eV for Fe \cite{LMA+06}. As double counting correction 
the around mean field expression (AMF)\cite{LAZ95,BCGN09,CS94} was used.
The Cr magnetic moments presented below are calculated within the MT
sphere with radius $R_{MT} = 2.32$ a.u. 
The experimental structure parameters used for CrSb$_2$ with marcasite
structure are: $a = 6.1481$ \AA, $b/a = 1.1404, c/a = 0.5428$ \cite{HK68a},
and for FeSb$_2$:  $a = 5.8328$ \AA, $b/a = 1.1208, c/a=0.5482$ \cite{HK68a}.

\section{Results}

Although there are several experimental results in the literature on
CrSb$_2$, no detailed theoretical investigation have been done
so far. Despite clear differences with FeSb$_2$, the two
systems have also some common features. Therefore, it is interesting to
analyze the properties of CrSb$_2$ in comparison with those
of FeSb$_2$.

\begin{figure}
\includegraphics[scale=0.3]{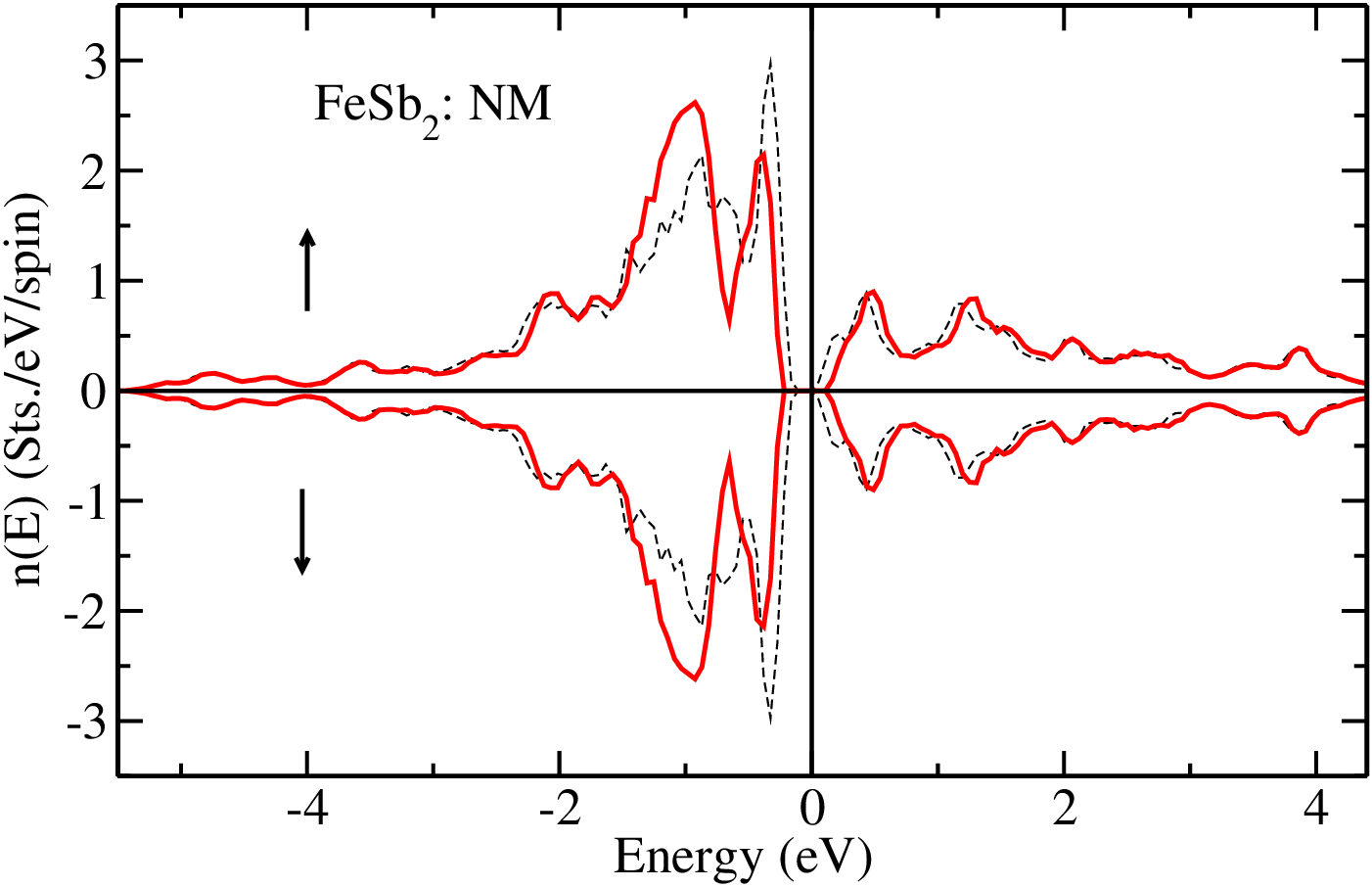} \;(a) \\
\includegraphics[scale=0.3]{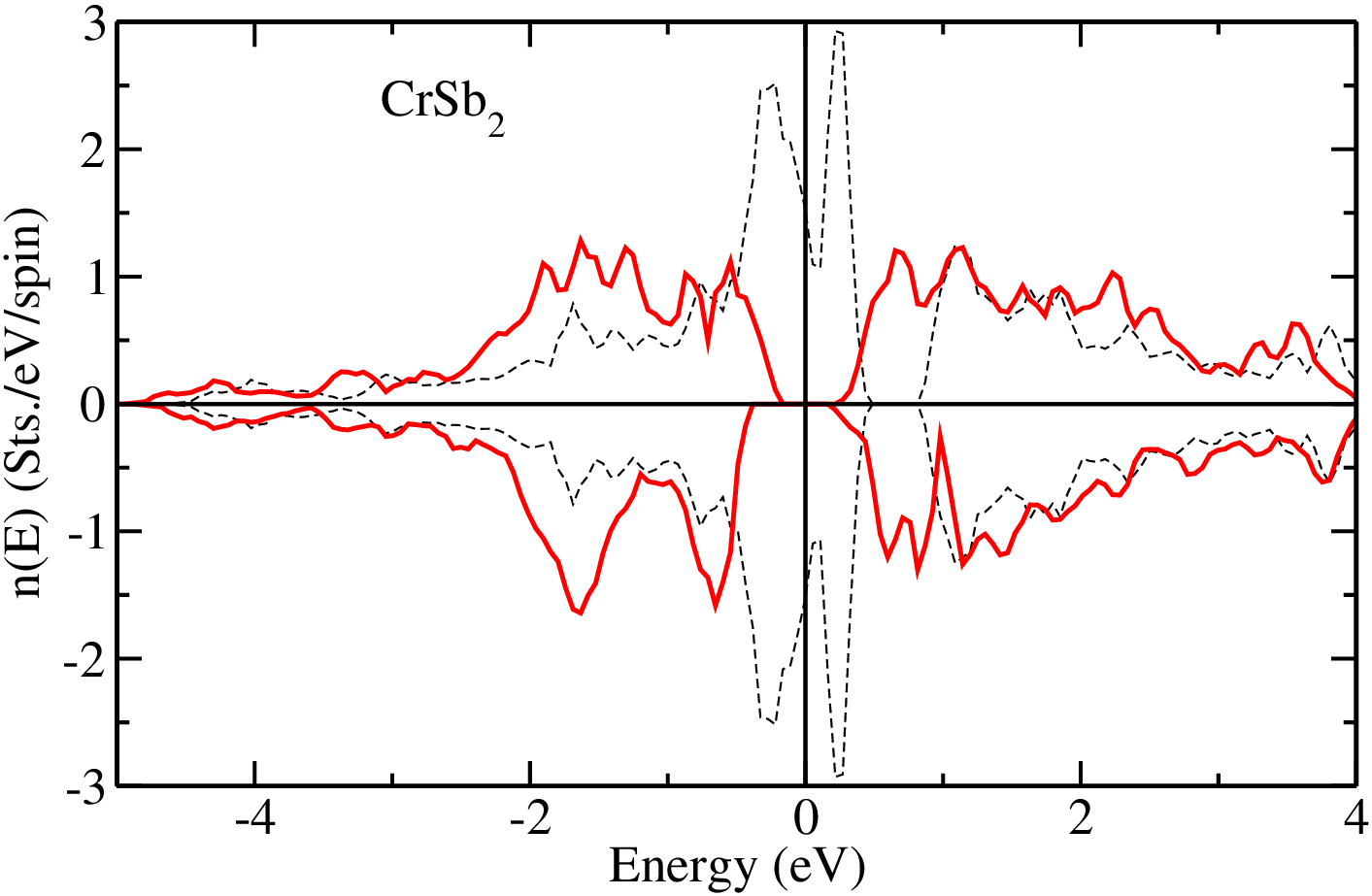}\; (b) \\
\includegraphics[scale=0.3]{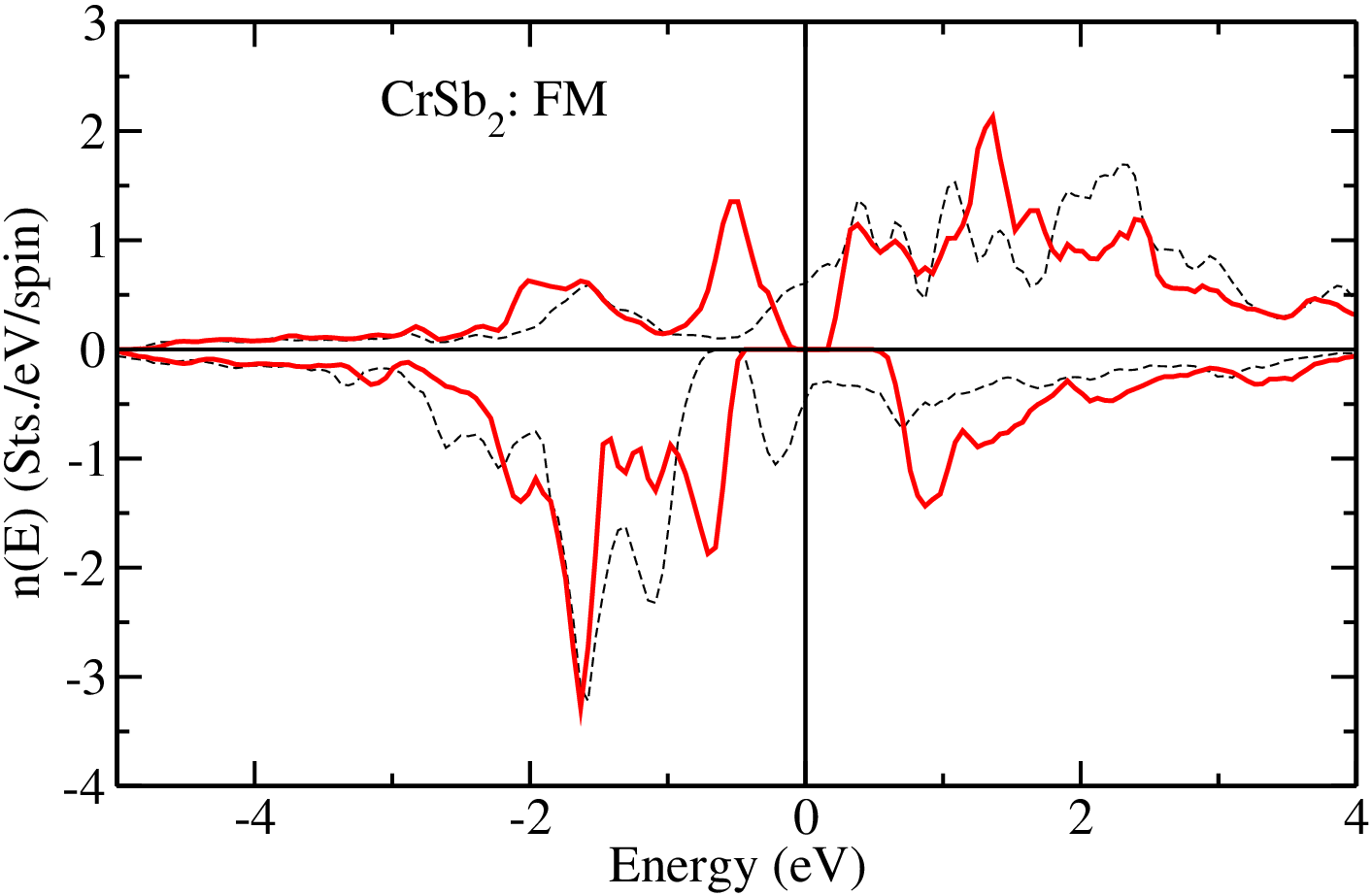}\; (c) \\
\includegraphics[scale=0.3]{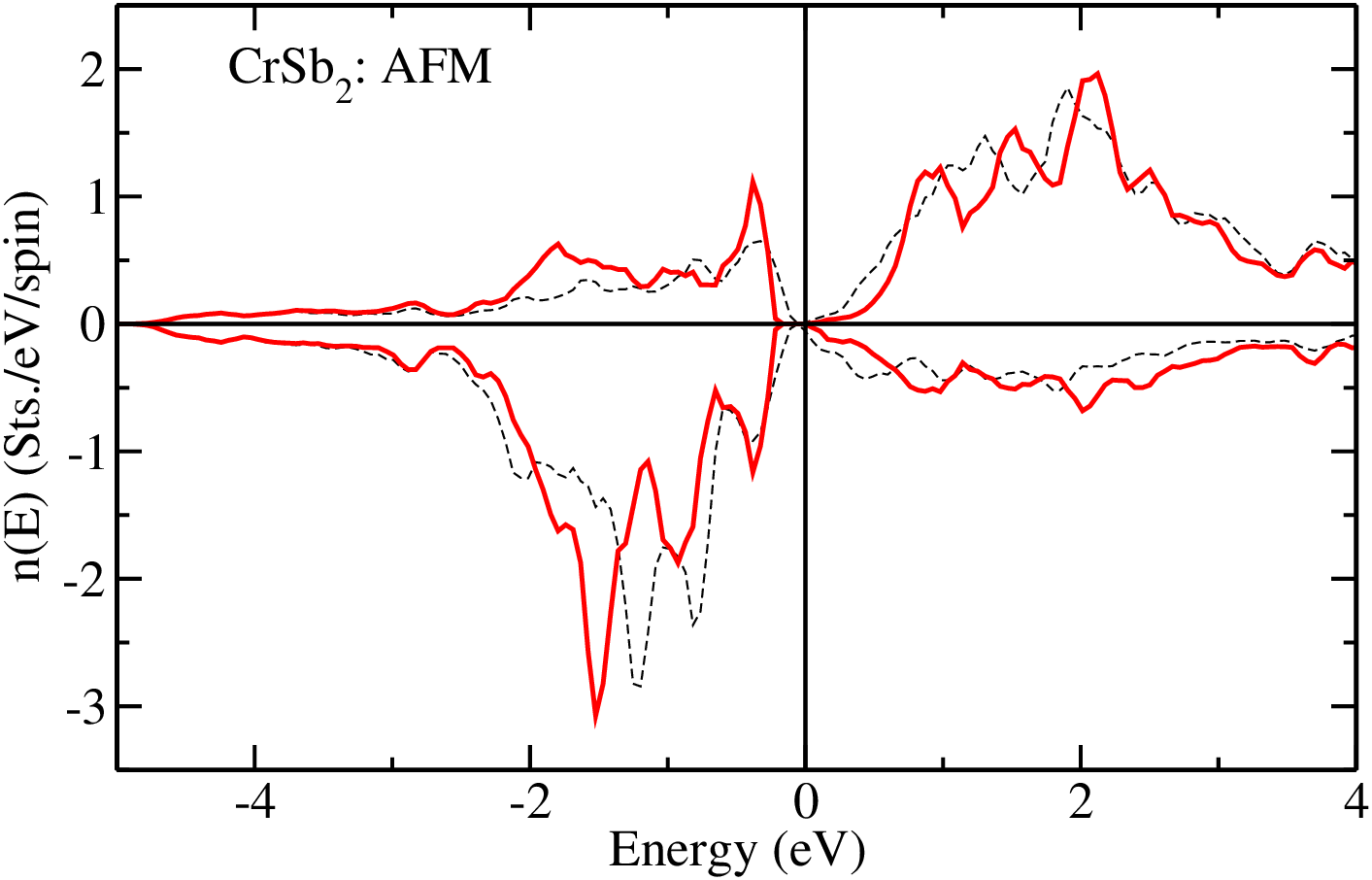}\; (d) \\
\caption{Density of states on the transition-metal atoms in FeSb$_2$ (NM 
  state) (a), and in CrSb$_2$ in NM (GGA results) and weakly magnetic with
  constrained magnetic moment $m_{Cr}=0.05 \mu_B$ (GGA+U results) (b), FM (c) and 
  AFM (d) states. Solid and thin dashed lines represent the results of GGA+U ($U
  = 2.7$ eV for Cr and  $2.6$ eV for Fe) and GGA calculations, respectively.}
\label{DOSCrSb2}
\end{figure}

\begin{figure}
\includegraphics[scale=0.3]{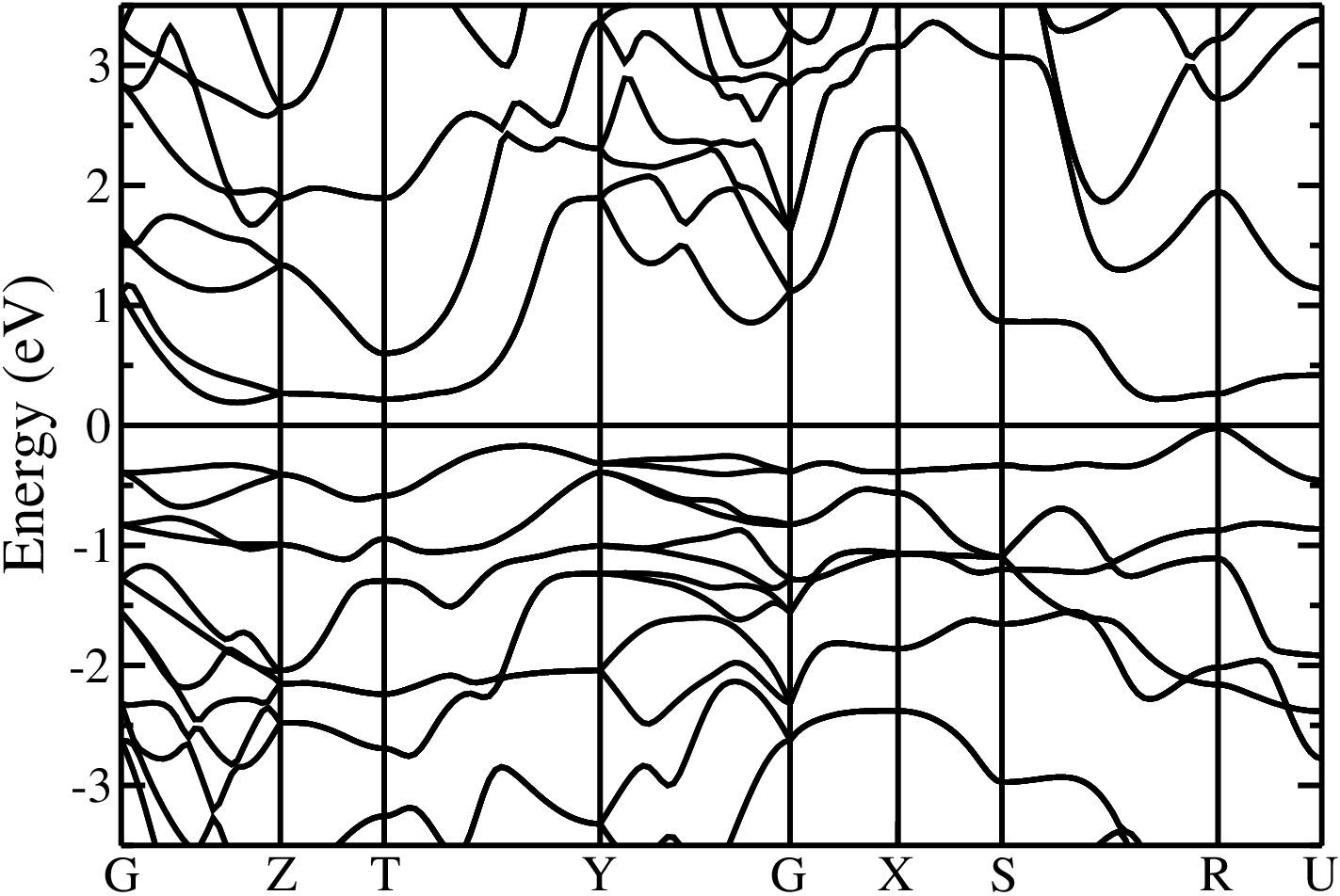} \;(a) \\
\includegraphics[scale=0.3]{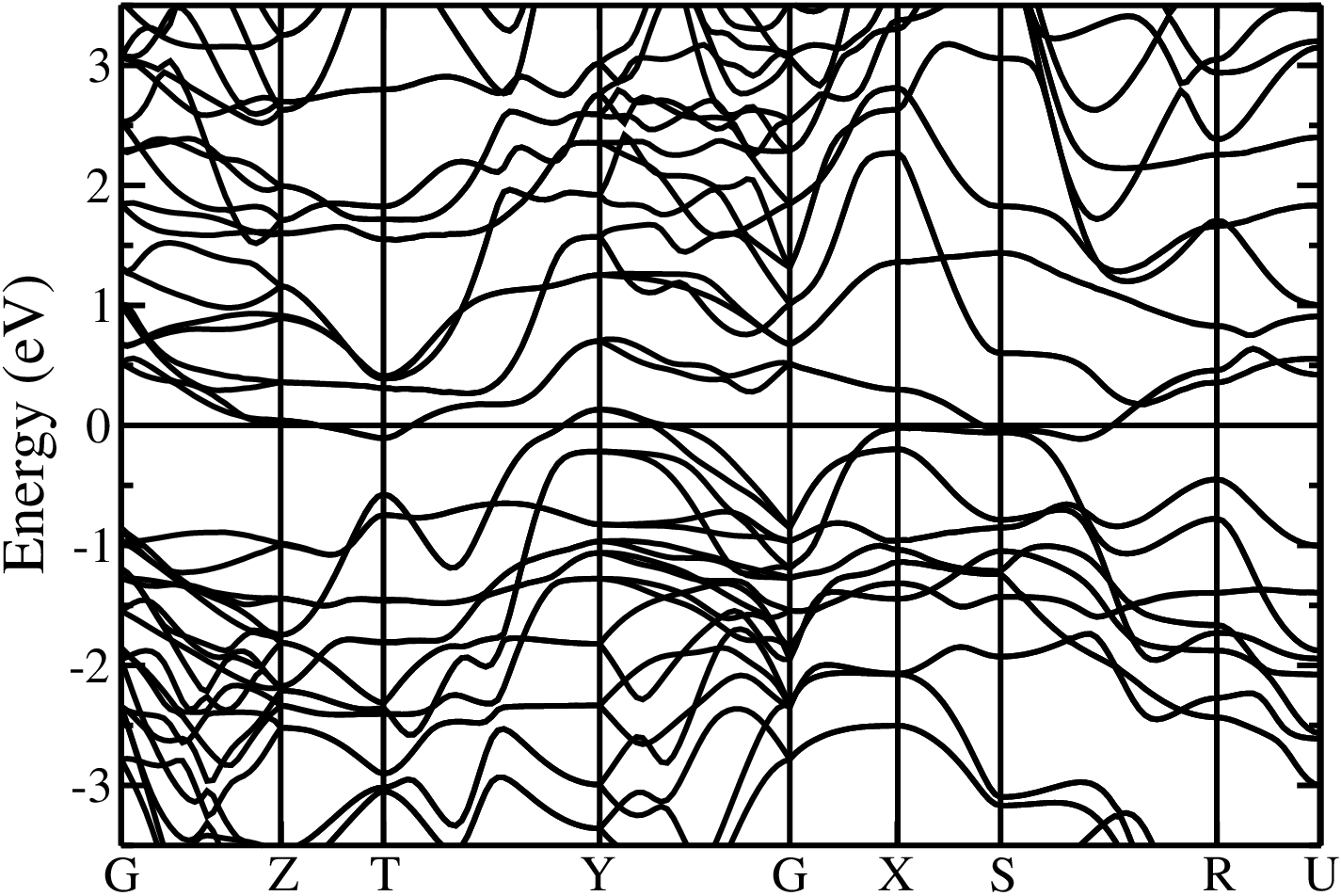} \;(b) \\
\includegraphics[scale=0.3]{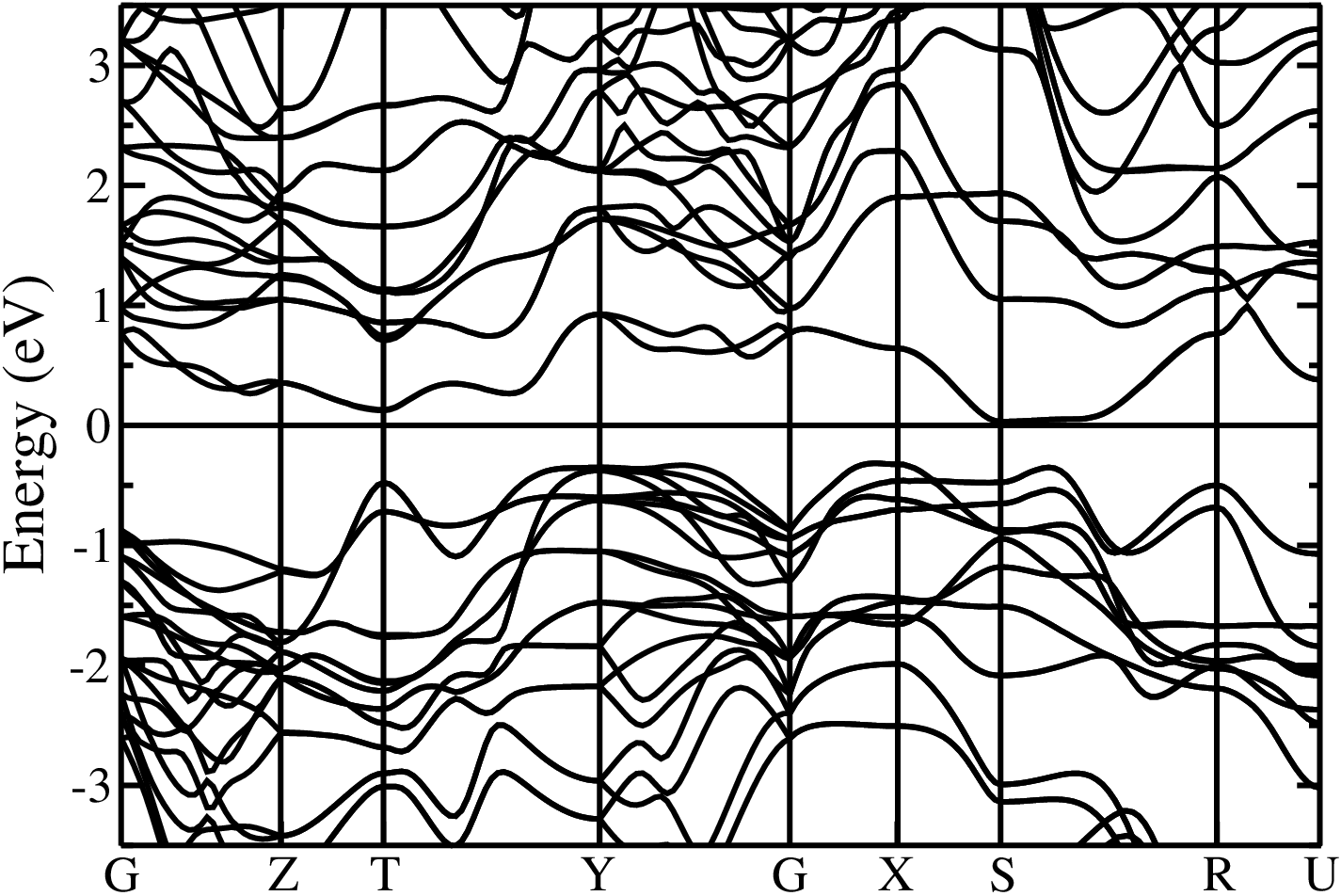} \;(c)
\caption{Electronic band structure for NM FeSb$_2$ via GGA   (a)
 and for FM CrSb$_2$  via GGA (b)  and  GGA+U, $U = 2.7$ eV (c)}.
\label{bandCrSb2FeSb2}
\end{figure}

As a first step, electronic structure calculations for FeSb$_2$  and CrSb$_2$ have
been performed based on the GGA using the experimental structure
parameters. In the case of CrSb$_2$ with a small unit cell these
calculations lead to a FM metallic ground state of the system (see the
corresponding density of states (DOS) plot
on the Cr atoms in Fig. \ref{DOSCrSb2}c), in contradiction to the
experimental results that show a small-gap semiconducting behavior for
the system with AFM order. 
To describe the AFM order observed experimentally, supercell calculations
are required, which will be discussed below. As a first step, we focus
on the FM state of the system.
In comparison, GGA calculations for FeSb$_2$ lead to a semiconducting
NM state with the Fe DOS shown in Fig.~\ref{DOSCrSb2}a. GGA+U
calculations performed to explain the finite temperature properties of
FeSb$_2$ as reported in the literature demonstrate 
the crucial importance of a proper description of
local correlations this compound. Motivated by this experience, GGA+U
calculations have been performed also for CrSb$_2$.


To investigate the dependence of the results on the $U$ parameter the
calculations have been performed for different $U$ value varying from
$0$ eV to $7$ eV (see, e.g., the discussions in \onlinecite{AEE+96}).
 In the following, if not otherwise noted, the results are presented for
$U = 2.7$ eV as this value gives a well defined semiconducting state for
CrSb$_2$ with FM order. However, it should be noted, that in the case of
AFM order, that is found experimentally \cite{HKA70,HMP07} 
for to the ground state, the $U$ 
value does not play a crucial role for the semiconducting properties. 

Fig. \ref{DOSCrSb2}b displays the DOS on the Cr sites for NM CrSb$_2$
exhibiting metallic behaviour for GGA calculations (dashed line) with
an energy gap above the Fermi level. 
Local correlations accounted for via GGA+U approach (with $U = 2.7$ eV) 
lead to a pronounced modification of the electronic structure even for  
a slightly spin-polarized system within LSM calculations with Cr
magnetic moment of $0.05 \mu_B$ (Fig. \ref{DOSCrSb2}b, solid line). This
results in a shift of the Fermi energy into the energy gap, i.e. the
system becomes semiconducting.

In the case of CrSb$_2$  with FM order, selfconsistent GGA calculations 
(see the DOS in Fig. \ref{DOSCrSb2}c) lead to a Cr magnetic moment of about $2.63
\mu_B$  and a 
metallic state of the system.  GGA+U results on the other hand exhibit the energy gap at the
Fermi energy, different for the two different spin channels, and a magnetic moment
of Cr atom of about  $1.64 \mu_B$.
Corresponding modifications of the electronic structure can be seen in the
energy band dispersion shown in Fig. \ref{bandCrSb2FeSb2}, b (GGA) and c
(GGA+U).

Calculations for the AFM state of CrSb$_2$ have been performed for 
the magnetic structure obtained experimentally \cite{HKA70}.
Interestingly, in this case a small energy gap at
the Fermi level occurs even within the GGA calculations. Local correlations
treated via GGA+U only enhance the energy gap (see
Fig. \ref{DOSCrSb2}d) but do not result in further noteworthy modifications of the
electronic structure.
 
\begin{figure}
\includegraphics[scale=0.3]{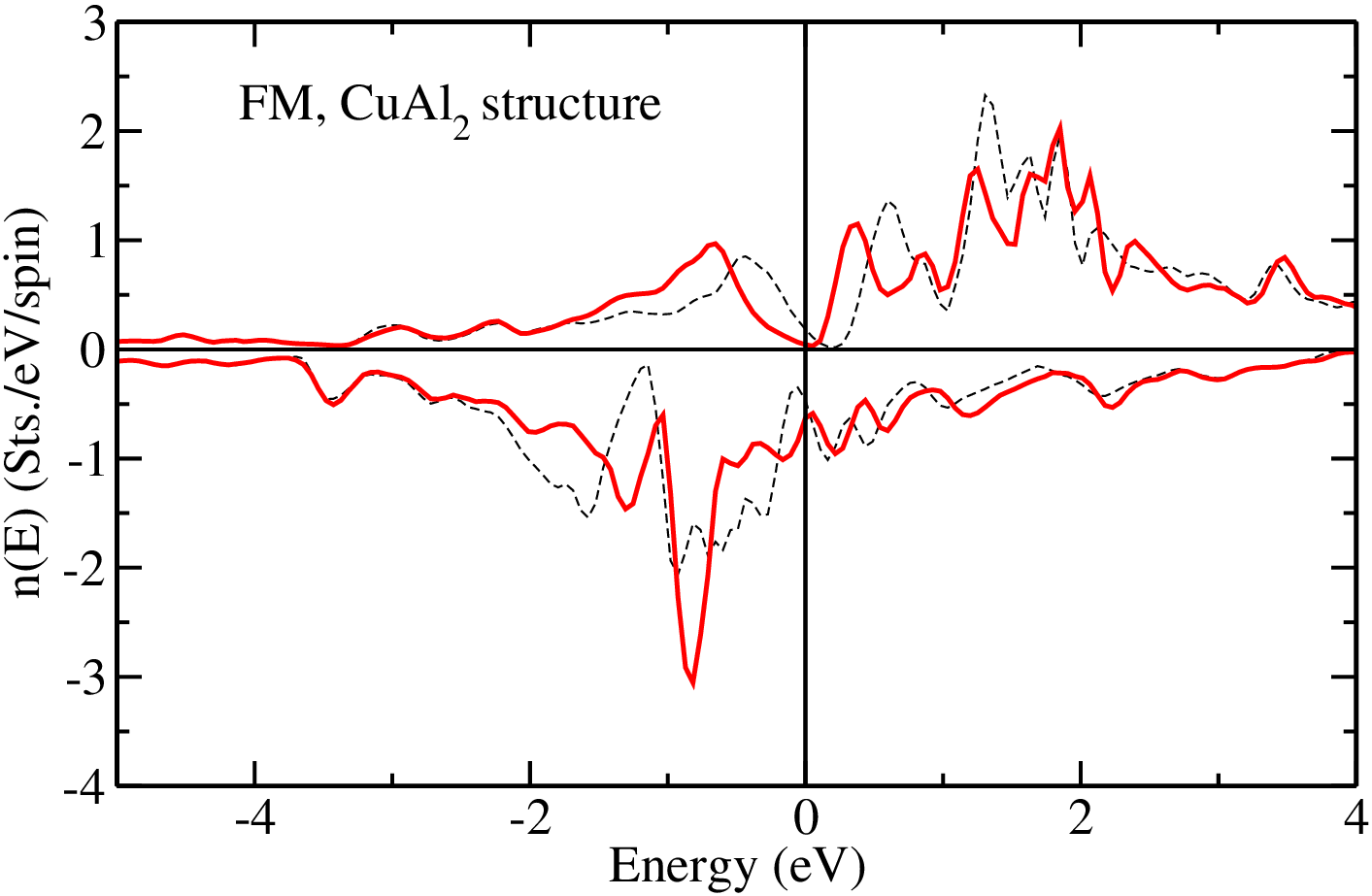} 
\caption{Density of states on the Cr atoms for CrSb$_2$
  with CuAl$_2$ structure in FM state. Solid and dashed lines represent
  the results of GGA+U ($U = 2.7$ eV) and GGA calculations, respectively.} 
\label{DOSCrSb2_2}
\end{figure}

\begin{figure}
\includegraphics[scale=0.3]{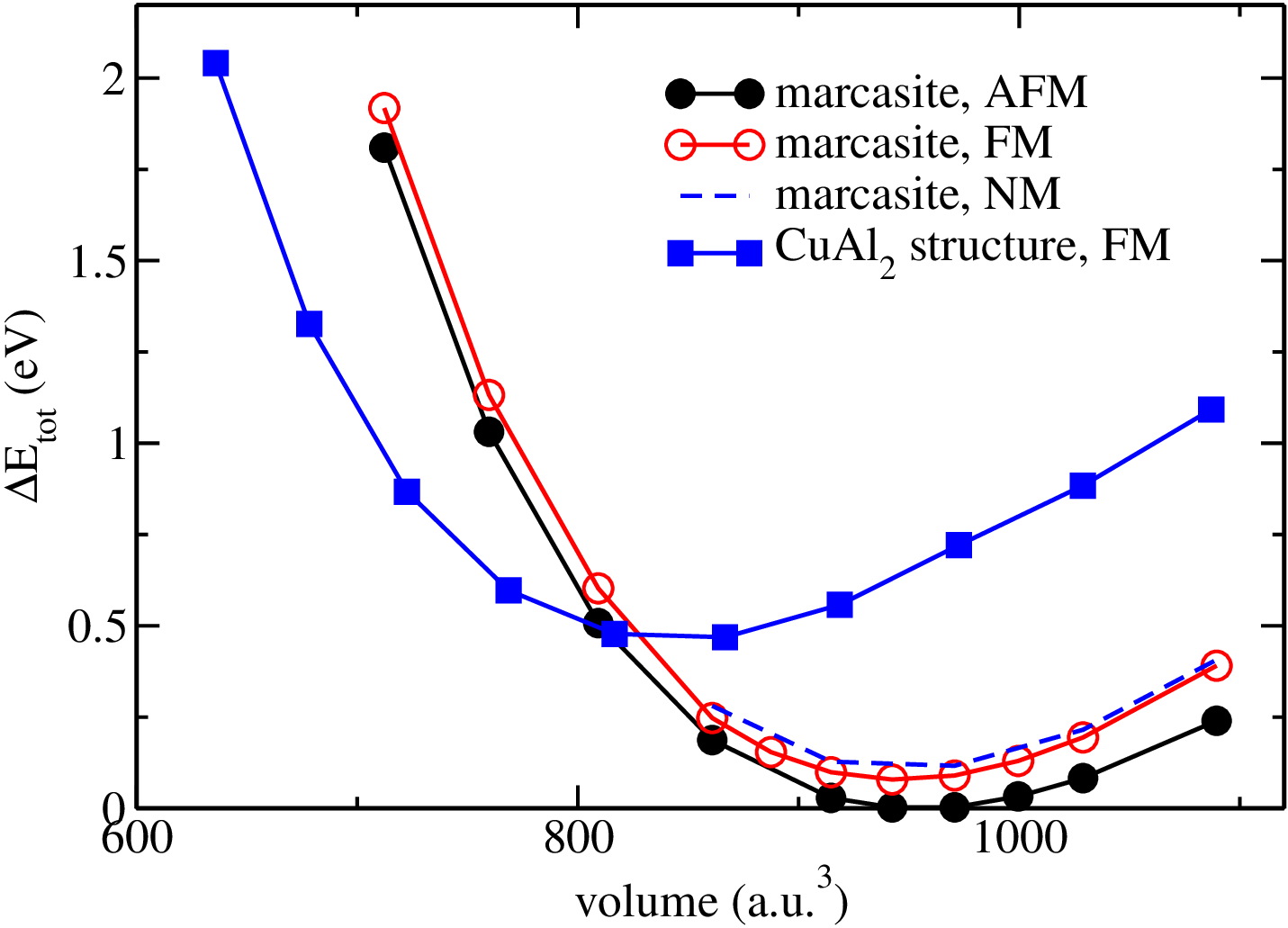}
\caption{Total energy of CrSb$_2$ calculated using the GGA+U for various
  magnetic phases with marcasite structure and for FM phase with
  CuAl$_2$ structure. The $c/a$ and $c/b$ ratios for all calculations have been
  kept constant at the experimental value \cite{HK68a,TUE99}.}
\label{etotCrSb2}
\end{figure}

Total energy calculations as a function of the volume have been
performed for the NM, FM and AFM states of CrSb$_2$ with marcasite structure
as well as for CrSb$_2$ in the FM state with CuAl$_2$ structure. The results
are presented in  Fig. \ref{etotCrSb2}. The volume in these
calculations have been varied by a variation of the lattice parameter $a$, 
keeping the ratios $b/a$ and  $c/a$ fixed as taken from experiment:
$b/a = 1.1404$ and $c/a = 0.5428$ for marcasite structure \cite{HK68a},
and $c/a = 0.885$ for CuAl$_2$ structure \cite{TUE99}.
All calculations have been performed within the GGA+U ($U = 2.7$ eV)
approach. Note that in the case of the CuAl$_2$ structure the difference in
DOS results obtained via GGA and GGA+U is rather small and in both cases
a metallic state of the system is observed (see Fig. \ref{DOSCrSb2_2}).

From these calculations one can see that the AFM state for CrSb$_2$ with
marcasite structure is indeed the ground state of the system as found in
experiment, while the FM and NM
states have slightly higher total energies. All total energy curves have a
minimum at the volume which is nearly the same for different types of
magnetic order and fits well to the experimental value. The curves for
the marcasite structure have a crossing point with the curve for
the CuAl$_2$ structure that has a minimum at a smaller volume than in
the case of marcasite structure. These results are fully in line with the
experimental data showing a transition from the marcasite to the FM
CuAl$_2$  phase at the pressure of 
about $5.5$ GPa \cite{TUE99}.

\begin{figure}
\includegraphics[scale=0.45]{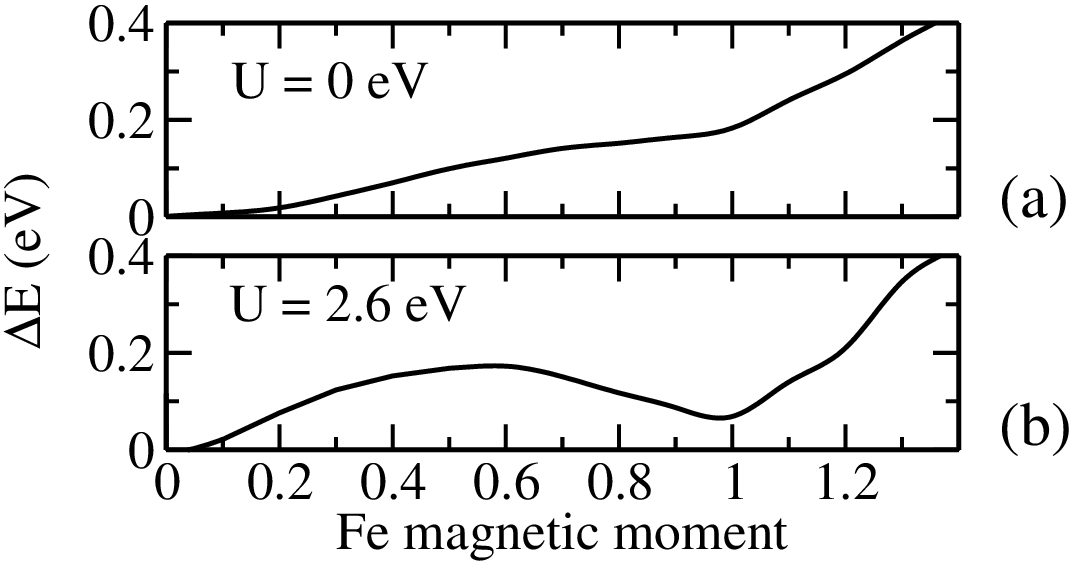}
\caption{The results of FSM calculations for FeSb$_2$: Total energy per
  unit cell vs fixed Fe magnetic moment via GGA (a) and GGA+U ($U = 2.6$ eV) (b).}
\label{FeSb_FSM}
\end{figure}

\begin{figure}
\includegraphics[scale=0.3]{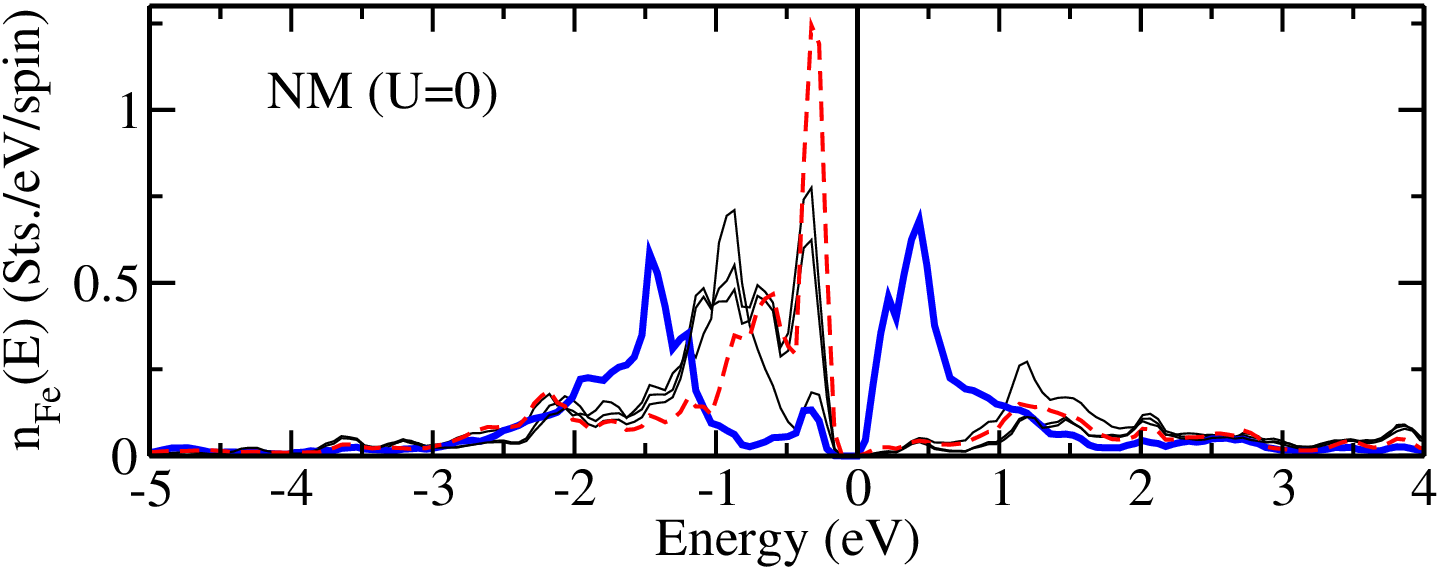} \;(a) \\
\includegraphics[scale=0.3]{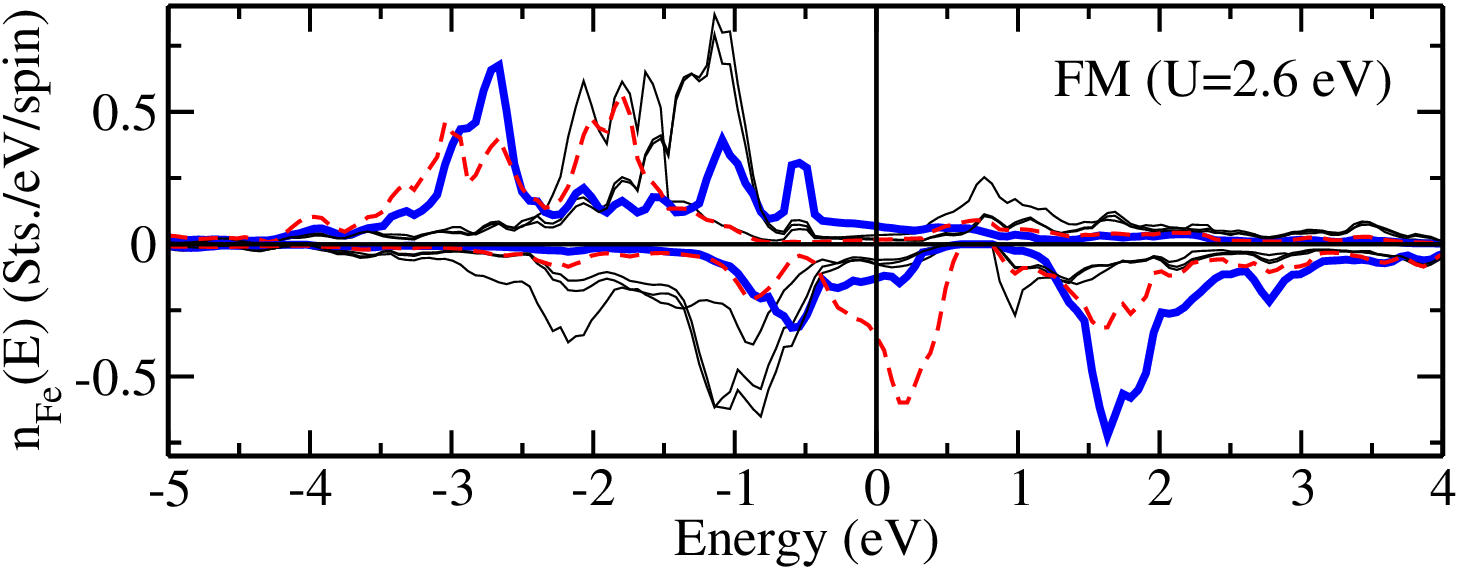} \;(b)
\caption{Fe DOS in FeSb$_2$ obtained via GGA+U: for the NM (a, $U = 0$ eV) and FM
  (b, $U = 2.6$ eV) states. Thick solid line corresponds to  $3d$-states with 
$d_{z^2}$ character, the dashed line shows the DOS of states with $d_{x^2-y^2}$ character.}
\label{FeSb_Fe_DOS}
\end{figure}

\begin{figure}
\includegraphics[scale=0.45]{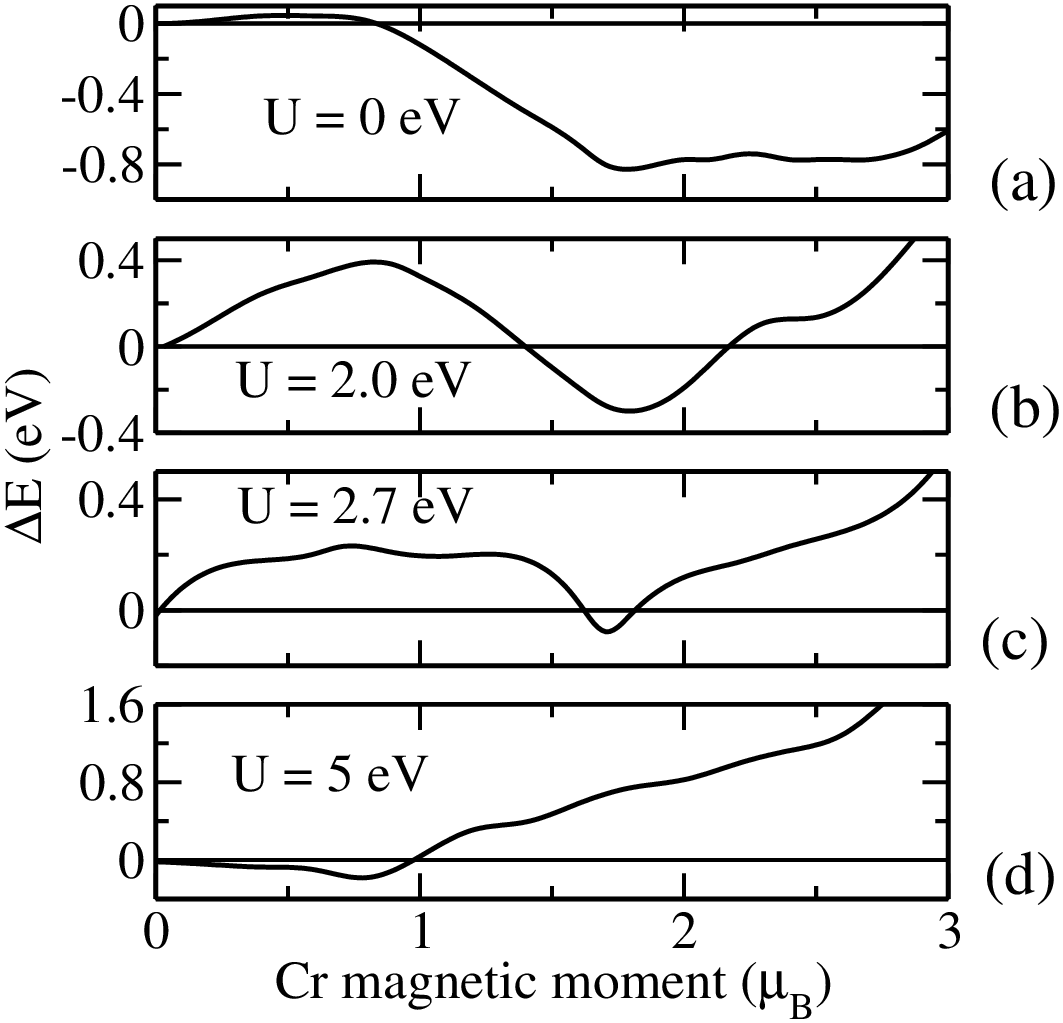}
\caption{The results of FSM calculations for  CrSb$_2$ with FM order.
  Total energy per unit cell vs fixed Cr magnetic moment via GGA (a)
  and GGA+U: $U = 2$ eV (b), $U = 2.7$ eV (c), $U = 5.0$ eV (d).}
\label{CrSb_FSM}
\end{figure}

\begin{figure}
\includegraphics[scale=0.3]{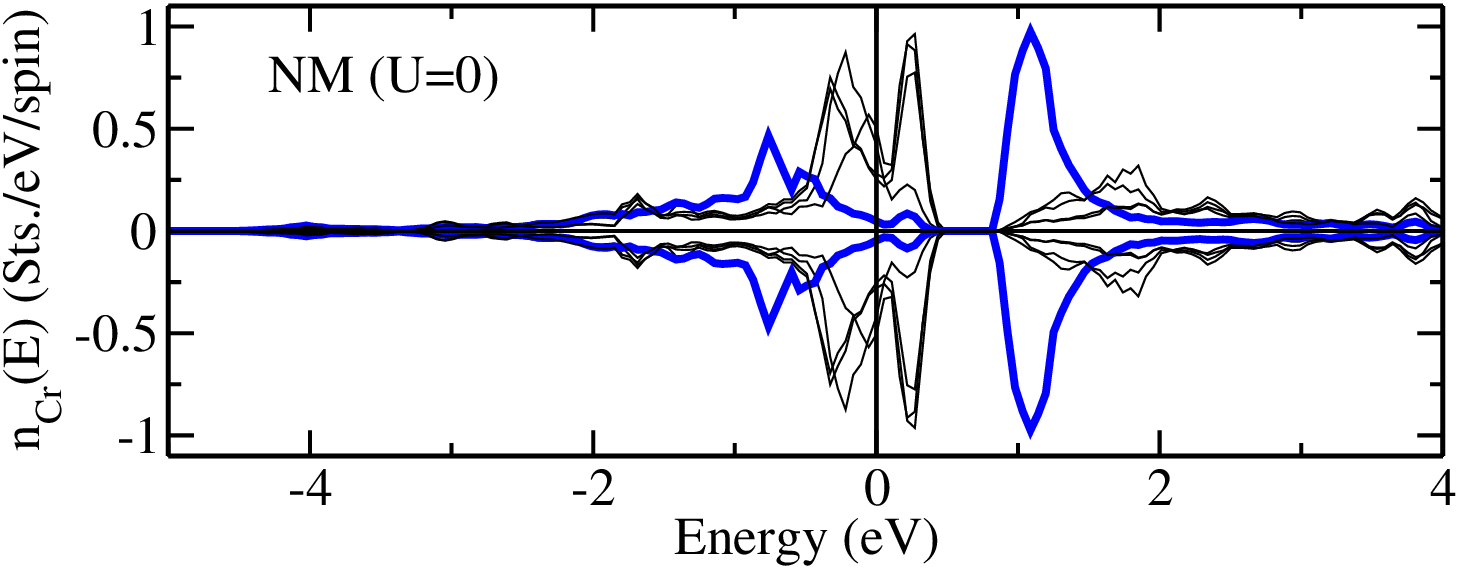} \;(a) \\
\includegraphics[scale=0.3]{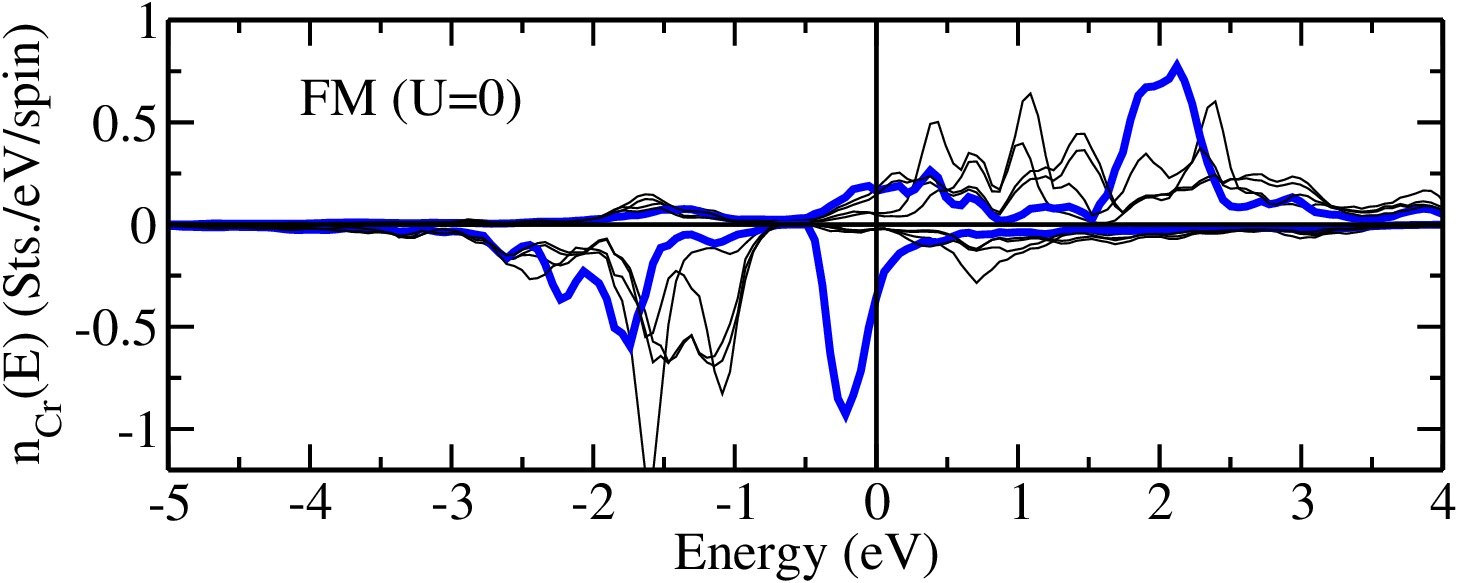} \;(b)
\includegraphics[scale=0.3]{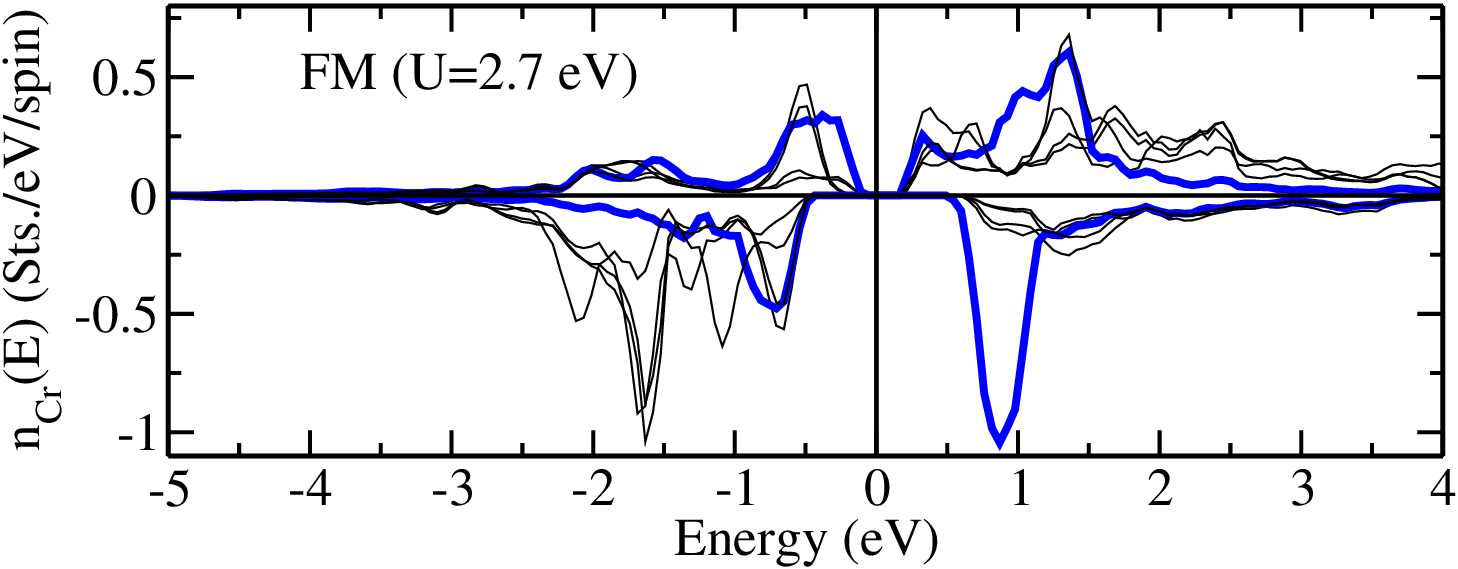} \;(c)\\
\includegraphics[scale=0.3]{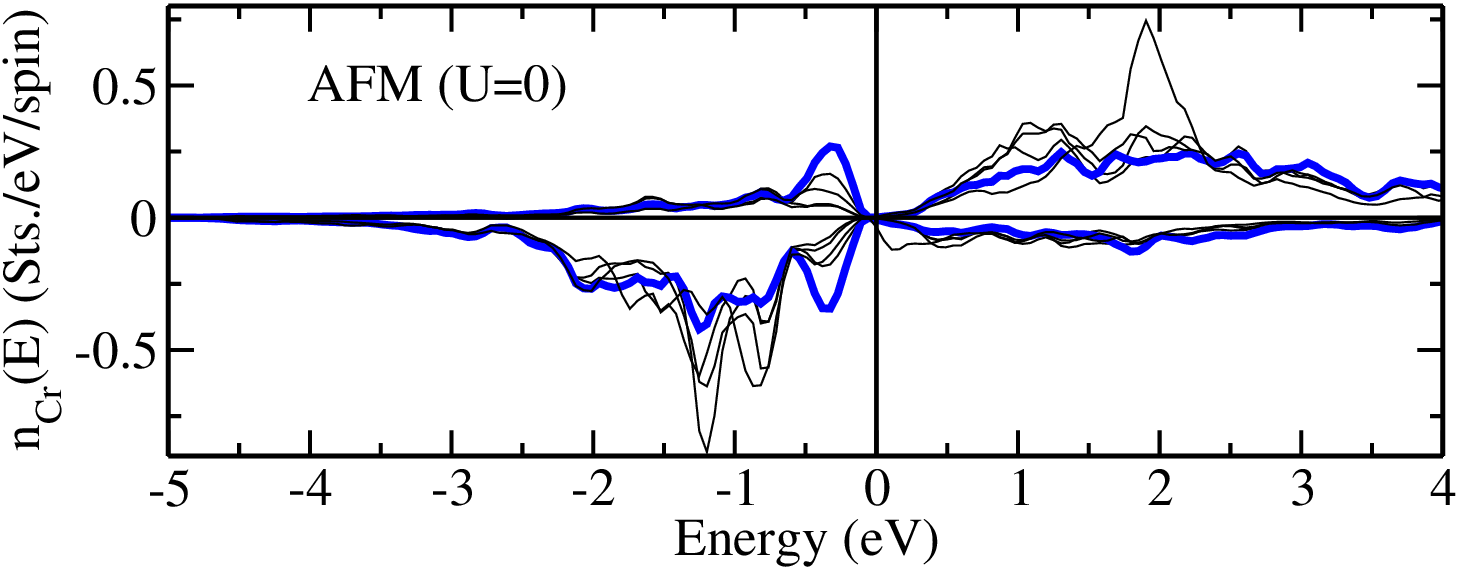} \;(d)
\includegraphics[scale=0.3]{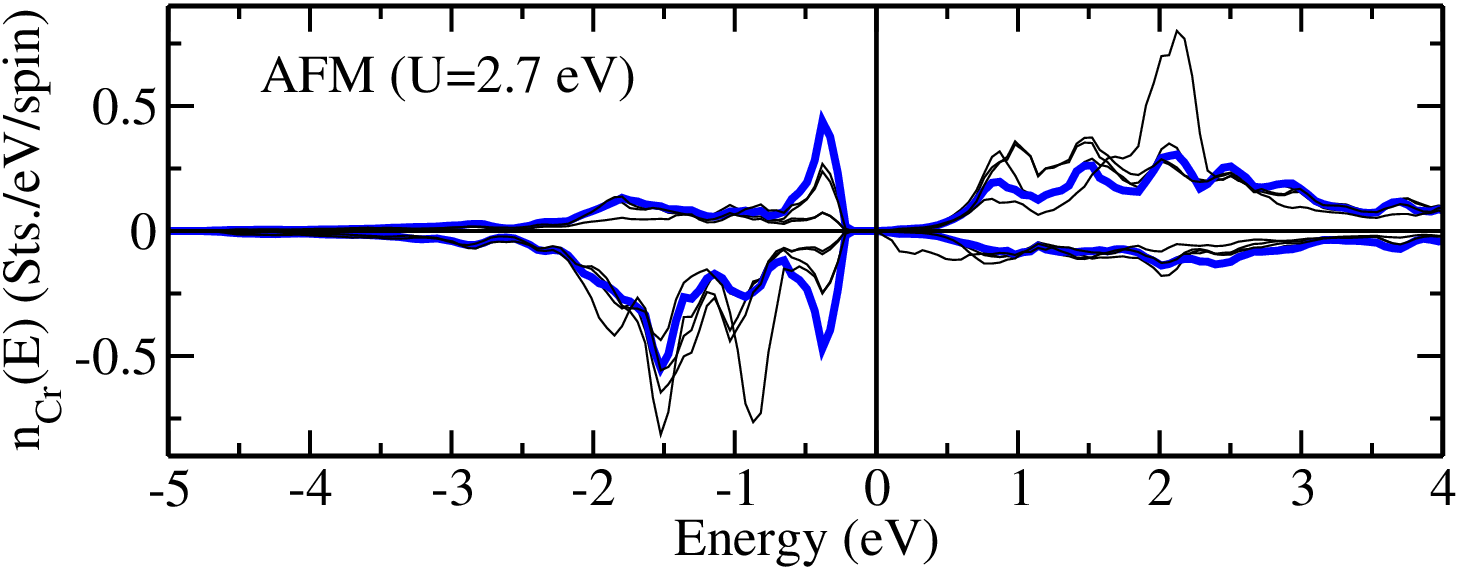} \;(e)
\caption{Cr DOS for CrSb$_2$ in  NM state (a, $U = 0.0$ eV),  FM state (b, $U =
  0.0$ eV) and (c, $U = 2.7$ eV), AFM state (d, $U = 0.0$ eV) and (e, $U
  = 2.7$ eV). Thick solid line corresponds to the $3d$-states with
$d_{z^2}$ character.}
\label{CrSb_Cr_DOS}
\end{figure}

\begin{figure}
\includegraphics[scale=0.3]{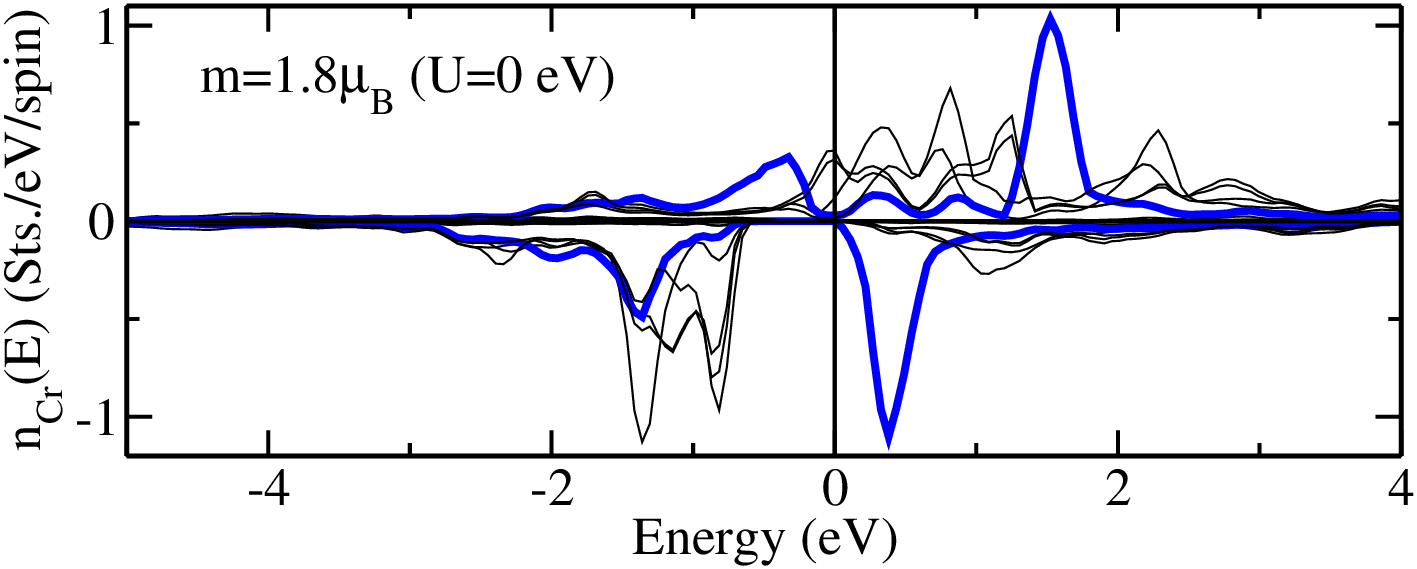} \;(a)
\includegraphics[scale=0.3]{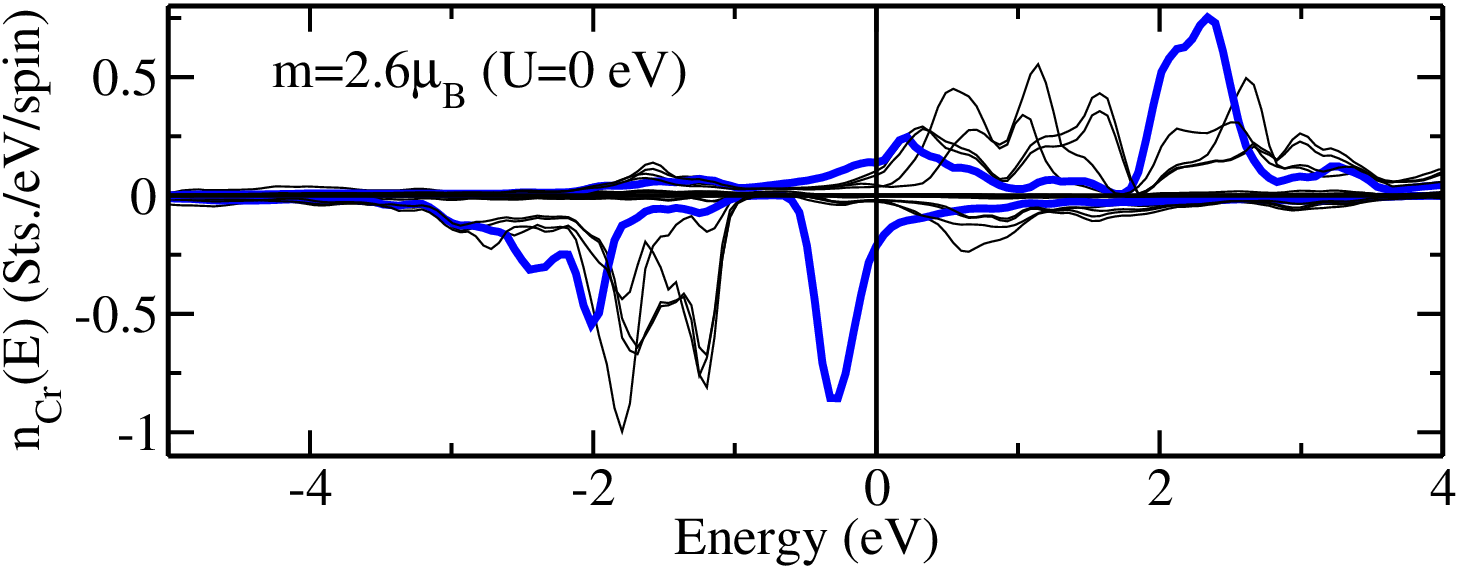} \;(b)\\
 \includegraphics[scale=0.3]{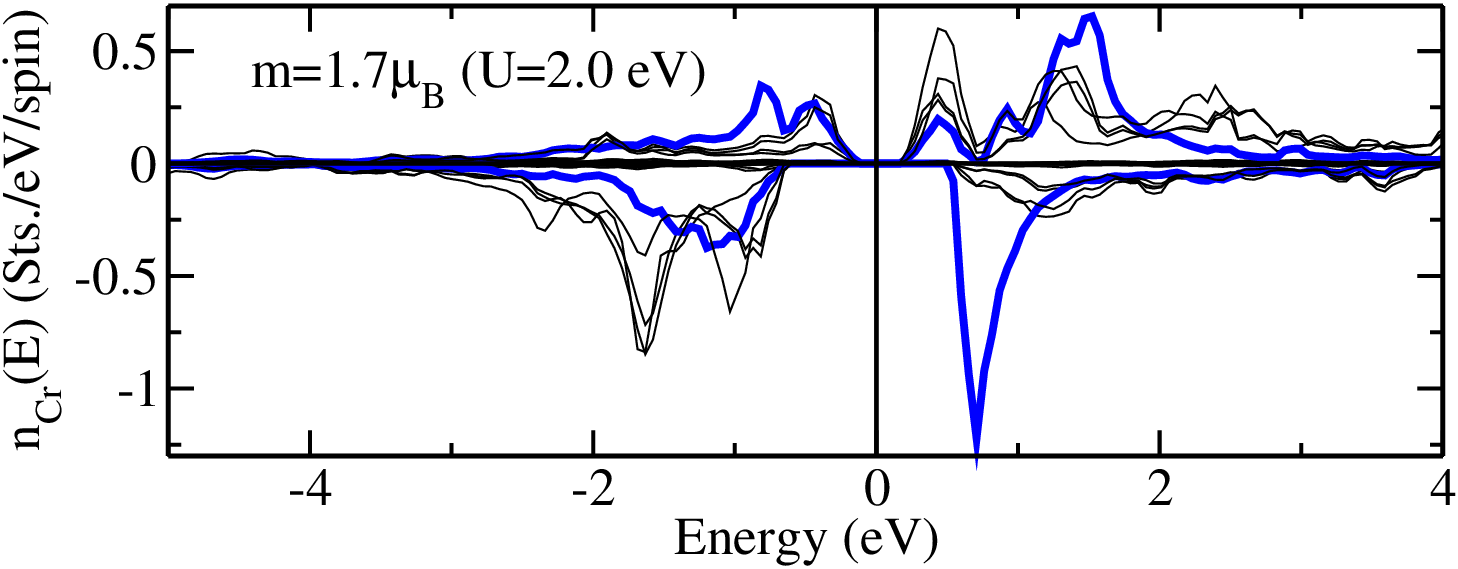} \;(c)
 \includegraphics[scale=0.3]{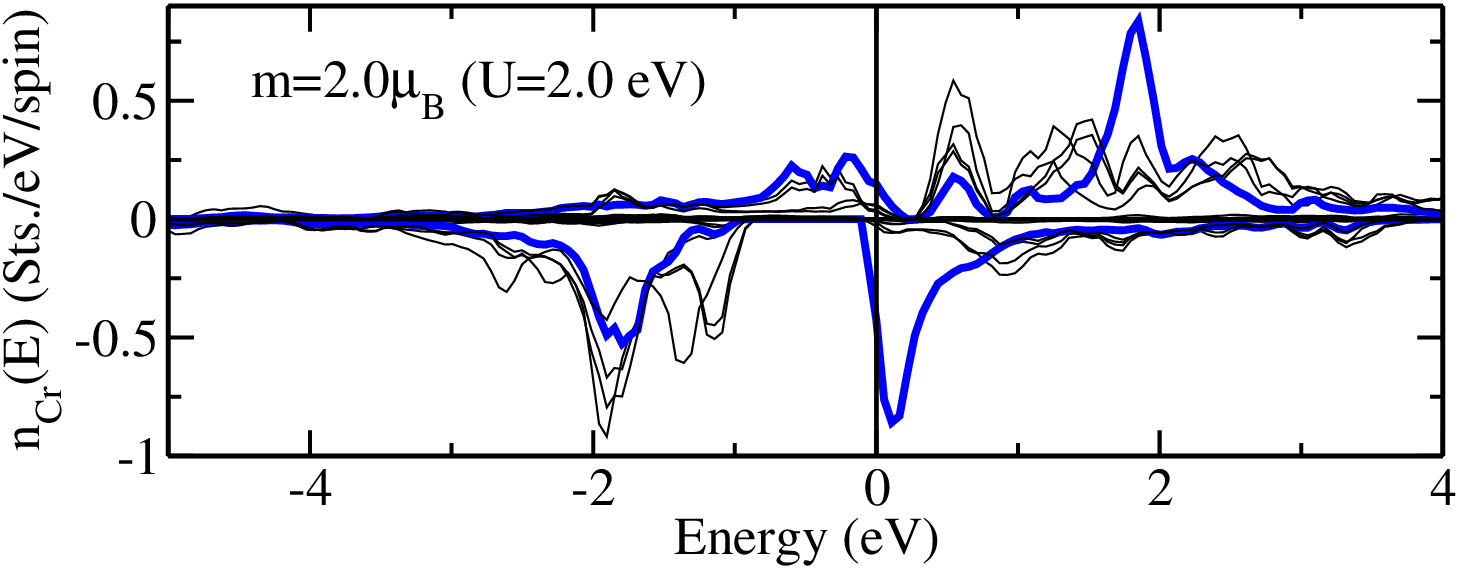} \;(d)\\
\caption{Results of FSM calculations for CrSb$_2$. Cr DOS in the FM state for
  the constrained Cr spin moment $m = 1.8 \mu_B$ (a) and $m = 2.6 \mu_B$
  (b) for $U = 0.0$ eV); and $m = 1.7 \mu_B$ (a) and $m = 2.0 \mu_B$
  (b) for $U = 2.0$ eV). Thick solid line corresponds to the $3d$-states with
$d_{z^2}$ character. }
\label{CrSb_Cr_DOS_FSM}
\end{figure}

The temperature dependence of
transport and magnetic properties of CrSb$_2$ and the influence of correlation effects can be analyzed  by
performing fixed spin moment (FSM) calculations and comparing the
results with the results for FeSb$_2$.
As was reported, e.g., by Lukoyanov et al. \cite{LMA+06}, the ground
state of  FeSb$_2$ is NM (see Fig. \ref{FeSb_FSM}a).
GGA+U calculations, on the other  hand, result in an additional minimum at $1
\mu_B$  (Fig. \ref{FeSb_FSM}b) that is used to explain the strong spin
fluctuations at high temperatures. According to the discussions by Lukoyanov
et al. \cite{LMA+06} on 
FeSb$_2$ as well as Anisimov et al. \cite{AEE+96} on FeSi, this
minimum is associated with the occupation of narrow Fe 3d-bands with 
$d_{z^2}$ character at the bottom of the conduction band, and states of $d_{x^2-y^2}$
character near the top of the valence band (see Fig. \ref{FeSb_Fe_DOS}).
%
%
The results of corresponding FSM calculations for CrSb$_2$ are shown in
Fig. \ref{CrSb_FSM}. They exhibit rather different features when
compared to FeSb$_2$.
The GGA calculations show an absolute minimum for the total energy 
around a Cr magnetic moment of $2.0 \mu_B$ (Fig. \ref{CrSb_FSM}a), in
contrast to FeSb$_2$ exhibiting a NM ground state.
This means, the NM state of CrSb$_2$ (with the Cr DOS shown in Fig. \ref{CrSb_Cr_DOS}a) is
unstable with respect to the creation of a local magnetic moment on the Cr atoms.
In general, this can lead either to FM or AFM order in the system with
the corresponding Cr DOS shown in Fig. \ref{CrSb_Cr_DOS}, b and d, respectively.
This minimum is split into a deeper at $\approx 1.8 \mu_B$ and a
more shallow one at $\approx 2.6 \mu_B$ which depends essentially on the
$U$ value.  
The Cr DOS curves corresponding to these two minima are plotted in
Fig. \ref{CrSb_Cr_DOS_FSM} a and b. One can see that the state with
$\approx 1.8 \mu_B$ is associated with the unoccupied majority-spin Cr
 $3d_{z^2}$-states, while magnetic moment $\approx 2.6 \mu_B$ is caused
 by additional occupation of Cr states of $d_{z^2}$ character.

In the  case of a higher magnetic moment the exchange splitting of
the electronic states leads to an occupation of antibonding
states in the spin-up channel and depleting of the bonding states in
the spin-down channel. Such an occupation leads to a partial compensation of
the energy gain due to the exchange splitting. Therefore, the FM state prefers to
have an unoccupied $3d_{z^2}$ spin-up energy band, that is stabilized by
treating  local correlation effects via the GGA+U approach (see the DOS 
for $U = 2$ eV in Fig. \ref{CrSb_Cr_DOS_FSM} c and d).
Thus, in the case of $U = 2 eV$ the total energy within the FSM
calculations has one minimum at $m = 0 \mu_B$ and a second broad minimum
in the regime $1.7$ to $2.0 \mu_B$. Fig. \ref{CrSb_Cr_DOS_FSM} c and d
shows the DOS corresponding to the 
values of Cr magnetic moment $m = 1.7 \mu_B$ and $m = 2.0 \mu_B$.
These results (for $U = 2$ eV, $2.7$ eV) exhibit a similar
behavior as in the case of FeSb$_2$ with the difference that the
minimum at the finite magnetic moment is deeper.
A further increase of the $U$ value above $5.$ eV results in an
instability of the FM state. 

 Another scenario is possible with a shift downwards in  energy of the
 majority-spin $3d_{z^2}$-states of Fe, changing their relative position
 with respect to the the $p$ states of Sb. This also leads to the
 arrangement of the Fermi level 
 within the energy gap. This scenario is associated with an AFM state with 
 the corresponding DOS shown in Fig. \ref{CrSb_Cr_DOS}d (for $U = 0$)
 and  Fig. \ref{CrSb_Cr_DOS}e 
 (for $U = 2.7$ eV). In this case GGA calculations result in a 
small-gap semiconducting state with a local Cr magnetic moment close to
that obtained for the FM state (see Table 1), since Cr spin-up $3d_{z^2}$-states  
in this case are also occupied. In contrast to FM case, the energy gap
at $E_F$ in this case occurs due to different spin-dependent
hybridisation of the Cr $d$ states with $p$ states of Sb atoms.
Also, in contrast to the FM state treated via GGA, AFM order in CrSb$_2$
is expected to be more stable since all bonding electronic states are
occupied and antibonding states are not occupied. Altogether the GGA+U
calculations result only in a slight variation of the electronic structure
leading in particular to a decrease of the Cr magnetic moments.

From the results above one can draw the following conclusions. 
Stabilization of the AFM order as a ground state of CrSb$_2$ is related 
to a shift below the Fermi energy of Fe spin-up states of $d_{z^2}$
character with a corresponding modification of the spin-dependent
hybridization of Cr $d$ states and $p$ states of Sb.
This moves the Fermi level into the energy gap to make 
all bonding states to be occupied and antibonding states unoccupied. 
 Local correlations treated via GGA+U influence in this case
 mainly the width of the energy gap. 
 In the case of the metastable FM state GGA+U calculations lead to more
 crucial modifications of the electronic structure resulting in a
 semiconducting state of the system. This occurs due to a push up of the
 majority-spin $d_{z^2}$ states of Cr above $E_F$ in cost of stronger
 hybridization of minority-spin states with $p$ states of Sb keeping
 the total number of electrons on the Fe atoms unchanged. 
 

Although the FSM calculations have been performed only for the FM order, 
the results above allow also to expect that being in the
AFM state, CrSb$_2$ should exhibit a metastable state determined by
a different occupation of the  $d_{z^2}$ Cr states.
Indeed, a temperature induced transition to the metastable state
could explain the behavior of the susceptibility and transport 
properties observed experimentally at low temperature.
It is possible that these two AFM states with slightly different
electronic structure can be characterized also by
different Cr local magnetic moments.

\begin{figure}
\begin{center}
\begin{tabular}{l|c|c|c}
              & GGA           & GGA+U     & Expt.       \\
\hline
marcasite AFM & 2.57          & 2.03      & 1.94      \\
\hline
marcasite FM  & 2.62          & 1.64      &  -      \\
\hline
CuAl$_2$      & 2.23          & 1.97      & 1.2 
\end{tabular}
\end{center}
\caption{Magnetic moments in $\mu_{Bohr}$ per atom for CrSb$_2$ (in
  the MT sphere with radius $R_{MT} = 2.32$ a.u.) within the GGA and
  GGA+U calculations. Theoretical results are compared with 
  experimental data.  \cite{HKA70,TUE99} }
\label{tabmom}
\end{figure}

\section{Conclusion}

In summary, one has to stress that the AFM order in CrSb$_2$ plays a
crucial role for the type of conductivity in the system exhibiting the
small-gap semiconducting properties. Although the local correlations
treated via the GGA+U approach influence mainly the width of the energy
gap of CrSb$_2$,
their role can be important to describe the experimental results as it
was shown for FeSb$_2$. In general, the present theoretical results
are in a good agreement with the experiment, showing the AFM ground
state of CrSb$_2$ with marcasite structure, with a Cr local magnetic
moment of about $2.03 \mu_B$ and with a small energy gap at the Fermi
level. Concluding from the total energy calculations a phase
transition under pressure to the CuAl$_2$ phase exhibiting the
properties of FM metal should be observed.

\begin{acknowledgments}

Financial support by the {\em Deutsche Forshungsgemeinschaft} within the
framework of the priority program {\em (DFG-Schwerpunktprogramm 1415)
 Kristalline Nichtgleichgewichtsphasen (KNG) - Pr\"aparation, Charakterisierung und in situ-Untersuchung der Bildungsmechanismen } is gratefully
acknowledged.

\end{acknowledgments}




\end{document}